\documentclass[]{aastex631}

\usepackage{graphicx}
\usepackage{epstopdf}

\begin{document}

\title{Observational Analysis of Multi-thermal Counter-streaming Flows in a Forming Filament and Their Relationship with Local Heating at Filament Footpoints}

\correspondingauthor{Hechao Chen}
\email{hechao.chen@ynu.edu.cn}

\author{Hongrui Li}
\affiliation{School of Physics and Astronomy, Yunnan University, Kunming 650500, People’s Republic of China}

\author[0000-0001-7866-4358]{Hechao Chen}
\affiliation{School of Physics and Astronomy, Yunnan University, Kunming 650500, People’s Republic of China}
\affiliation{Yunnan Key Laboratory of Solar Physics and Space Science, Kunming, 650216, People’s Republic of China}

\author[0000-0002-7153-4304]{Chun Xia}
\affiliation{School of Physics and Astronomy, Yunnan University, Kunming 650500, People’s Republic of China}
\affiliation{Yunnan Key Laboratory of Solar Physics and Space Science, Kunming, 650216, People’s Republic of China}

\author{Xiaoli Yan}
\affiliation{Yunnan Observatories, Chinese Academy of Sciences, Kunming 650216, People’s Republic of China}
\affiliation{Yunnan Key Laboratory of Solar Physics and Space Science, Kunming, 650216, People’s Republic of China}

\author{Yongyuan Xiang}
\affiliation{Yunnan Observatories, Chinese Academy of Sciences, Kunming 650216, People’s Republic of China}


\author{Jincheng Wang}
\affiliation{Yunnan Observatories, Chinese Academy of Sciences, Kunming 650216, People’s Republic of China}
\affiliation{Yunnan Key Laboratory of Solar Physics and Space Science, Kunming, 650216, People’s Republic of China}




\begin{abstract}
Utilizing high-resolution imaging and spectroscopic observations from the New Vacuum Solar Telescope (NVST), the Interface Region Imaging Spectrograph (IRIS), and the Solar Dynamics Observatory (SDO), we investigated the nature and origin of counter-streaming flows within a forming active region filament. The ever-present counter-streaming flows observed within the filament are identified as interleaved unidirectional mass flows in opposing directions occurring in neighboring threads. Multi-wavelength observations corroborate the multi-thermal nature of these counter-streaming flows: the cool H$\alpha$ component flows at about 10$-$20 km s$^{-1}$ , while the warm ultraviolet and extreme ultraviolet components reach 40–70 km s$^{-1}$. The Si~{\sc{iv}} 1400 Å line reveals significant micro-turbulence in the filament's counter-streaming flows, with a nonthermal velocity width of $\sim$40  km s$^{-1}$.
These multi-thermal flows emanate from compact brightenings at the filament’s ends, manifesting as small-scale, collimated upflows at their nascent phase. They continuously inject both chromospheric and transition region plasma into the filament channel, thereby feeding the counter-streaming flows. At their base,  the Si~{\sc{iv}} and C~{\sc{ii}} spectral lines show pronounced line broadening and intensity enhancements, indicating significant localized chromospheric heating. Additionally, numerous small-scale photospheric flux emergence and cancellation events, with  magnitude of $\sim$  $10^{17}$ Mx, are detected near their base. We suggest such weak magnetic-field activities, possibly associated with unresolved magnetic reconnection events, drive these persistent upflows and localized footpoint heating. This work elucidates the multi-thermal origin of counter-streaming flows within a forming filament and provides evidence of localized chromospheric heating at the filament footpoints. 
\end{abstract}

\keywords{Solar physics (1476)---Solar filaments (1495)--- Solar magnetic fields (1503) --- Solar chromosphere (1479) --- Jets (870)}

\section{Introduction} \label{sec:intro}

Solar filaments, also known as prominences when appearing beyond the solar limb, are elongated dark filamentary  plasma structures supported by special magnetic field topologies against the solar disk \citep{1995ASSL..199.....T}. They are always observed above the polarity inversion lines (PILs) of the magnetic field at the photosphere. Historically, they are classified by their evolutionary state of their associated magnetic fields into active-region (AR), quiescent, and intermediate types  \citep{2014LRSP...11....1P}. Most filaments observed in H$\alpha$, it is possible to identify a narrow filament main body, known as spine, running horizontally above the PIL. Long-lived quiescent filaments usually exhibit lateral extensions diverging from the filament spine, known as barbs, though these are not a ubiquitous feature for short-lived AR filaments \citep[see the review of][]{2015ASSL..415.....V}. Earlier views proposed that barbs terminate at the chromosphere, and possibly even reach the photosphere, representing the magnetic-field footpoints of the filaments intersecting the solar surface \citep[i.e.,][]{1998SoPh..182..107M}. However, high-resolution observations suggested that barbs bend down to enter and exit the chromosphere, just close to the photosphere \citep[e.g.,][]{2005SoPh..227..283L,2005SoPh..226..239L,2005ApJ...626..574C,2013SoPh..282..147L,2015RAA....15.1725Y}. 
Typical AR filaments have footpoints only at their extreme ends, whereas longer filaments in ARs or the quiet Sun may possess additional, intermediate footpoints \citep{2015ASSL..415.....V}.

It is widely accepted that filament material originates from the chromosphere, given the similarity in elemental composition and temperature between filament material and the chromosphere \citep{1998ApJ...494..450S,2017ApJ...836L..11S}. To date, several possible mechanisms are attributed to the origin of filament mass: direct chromospheric mass injection \citep[e.g.,][]{1999ApJ...520L..71W,2000SoPh..195..333C,2018ApJ...863..180W,2019MNRAS.488.3794W}, the evaporation-condensation process \citep{1991ApJ...378..372A,2012ApJ...758L..37B,2012ApJ...748L..26X,2012ApJ...745L..21L,2020ApJ...905...15V,2021ApJ...921L..33Y,2022A&A...659A.107C}, and the direct levitation of dipped field lines from the lower chromosphere \citep{1994SoPh..155...69R,2005ApJ...622.1275L,2008ApJ...673L.215O}, as well as the lifting mechanism of reconnected magnetic fields \citep[e.g.,][]{1999SoPh..190...45L,2005ApJ...630..587L}. 
Recent numerical simulations indicate that as the local heating site ascends, the direct chromospheric mass injection can convert to an evaporation-condensation process \citep{2021ApJ...913L...8H}. 

High-resolution {H$\alpha$} observations have revealed that all types of filaments consist of numerous thin and dark thread-like plasma structures when seen against the solar disk. Filaments, as a whole, persist as stable structures for hours to days, but individual threads are short-lived (lifetime $\sim$ minutes) and highly dynamic. 
Each of these thin threads exhibits pervasive and continuous mass flows. \citet{1991A&A...252..353S}  reported the presence of both blueshifted and redshifted motions within these threads, caused by bidirectional/antiparallel flows. \citet{1998Natur.396..440Z} reported bidirectional/antiparallel flows of 5$-$20 km s$^{-1}$ throughout the entire filament structure in both the red and blue wings of the H$\alpha$ line, which they first termed filament counter-streaming flows. Since then, the ubiquitous existence of similar counter-streaming flows  in different filaments have been extensively confirmed \citep{2003SoPh..216..109L,2005SoPh..226..239L,2007AdSpR..39.1700C,2008SoPh..247..321S,2008ASPC..383..243P,2010ApJ...721...74A,2018A&A...611A..64D}. Using ultra-high resolution observations from the High-resolution Coronal Imager, \citet{2013ApJ...775L..32A} first reported the existence of extreme ultraviolet (EUV) counter-streaming flows along AR filament threads within the million-degree corona,  with velocities of 70–80 km s$^{-1}$. The measured velocity of EUV counter-streaming flows is approximately three to four times higher than that of H$\alpha$ counter-streaming flows. 

Counter-streaming flows are now known to be commonplace within all filaments, but the nature and driving mechanism of these flows remain a subject of debate \citep[See the review of ][]{2020RAA....20..166C}.
In the past, counter-streaming flows have been interpreted as longitudinal oscillations within filament threads, where plasma moves back and forth in gravitational potential wells along the same magnetic field line with differing phases \citep{2005SoPh..226..239L}. Instead, other researchers believed that counter-streaming flows correspond to unidirectional flows in adjacent threads, where plasma moves along distinct magnetic field lines in opposite directions \citep{2015RAA....15.1725Y,2016ApJ...831..123Z,2018ApJ...852L..18W}. Numerous mechanisms have been proposed to explain the origin of such unidirectional flows, including siphonic flows driven by the gas pressure imbalance along flux tubes \citep{2014ApJ...784...50C}, opposite oscillation mass flows along distinct field lines \citep{2003SoPh..216..109L}, and small-scale energy releasing events such as weak EUV/UV brightenings \citep{2016A&A...589A.114L,2019ApJ...881L..25Y} or recurrent network jets  \citep{2020ApJ...897L...2P} at magnetic field footpoints. In fact, counterstreaming flows in different parts of a single prominence/filament often demonstrate quite complex mass motion patterns. 
For instance, \citet{2015ApJ...814L..17S} reported that counterstreaming flows in the horizontal prominence foot appear to be driven by the imbalanced gas pressure at both ends, while counterstreaming flows in the vertical prominence foot resulted from the coexistence of continuous upward and downward mass flows due to the collapse of prominence bubbles. 
\citet{2025ApJ...981..139Y} recently found that in the upper section of a limb quiescent solar prominence, large-amplitude and long-period longitudinal mass oscillations caused counterstreaming flows, while smaller-amplitude random oscillations and mass injection dominated its middle and lower sections. \citet{2014ApJ...784...50C} suggested that counterstreaming flows may result from the combined effect of one or even several of the mechanisms mentioned above, especially the combination of longitudinal oscillations and unidirectional flows in adjacent threads. Indeed, magnetohydrodynamic (MHD) simulation results from \citet{2020NatAs...4..994Z} suggest that if turbulent heating occurs at the footpoints of filament field lines, random evaporation of material would drive hot unidirectional flows and form cool filament threads. meanwhile, gas pressure driven by turbulent heating can also lead to subsequent oscillations of these threads with random phases, resulting in apparent cool counterstreaming flows.

In this study, we investigate counterstreaming flows in a forming active region (AR) filament using observations from the New Vacuum Solar Telescope (NVST), the Solar Dynamics Observatory (SDO), and the Interface Region Imaging Spectrograph (IRIS). This case study allows us to further quest the multi-thermal nature and possible origin of counterstreaming flows in a forming AR filament, utilizing high spatiotemporal resolution observations spanning the chromosphere, transition region, and corona. 

The remaining paper is organized as follows: In Section 2, we systematically introduced the observation instruments used in the research institute and the multi-band data obtained. Section 3 Based on the processed data, multi-band analysis and spectral diagnosis were carried out to reveal the physical connection between the counterstreaming flows and the localized heating at the filament's ends. Section 4 summarizes the key points of this work and discusses their implications.

\section{Instruments and Data Analysis} \label{sec:style}
The forming filament under investigation originated within a significantly decayed AR NOAA 12782  located in the southern hemisphere. The Global Oscillation Network Group \citep[GONG,][]{1996Sci...272.1284H} continuously monitored the formation and evolution of this filament in H$\alpha$. On November 14, this filament was well observed by the New Vacuum Solar Telescope  \citep[NVST,][]{2014RAA....14..705L} between 01:25 and 05:45 UT and by the Interface Region Imaging Spectrograph \citep[IRIS,][]{2014SoPh..289.2733D} from 02:45 to 03:34 UT on November 14. This temporal overlap in observations provides us with a valuable opportunity to investigate the mass flows and their origin within the forming filament.

We used the NVST H$\alpha$ line-center and the H$\alpha$ off-band images acquired at 6562.8 \AA, 6562.8 $\pm$ 0.6 \AA\ to study the dynamic characteristics and track the filament footpoints. The spatial resolution of these images is approximately  0$^{\prime\prime}$.165 \text{pixel}$^{-1}$, with a temporal cadence of 45 s \citep{2014RAA....14..705L}. To better reveal the mass flow pattern in this forming filament, Doppler proxy maps were constructed using the NVST H$\alpha$ red and blue wing data following the equation: $D = (I_B - I_R)/(I_R + I_B)$, where $I_R$ and $I_B$ represent the absolute pixel intensity observed at the red and blue wings, respectively. All the NVST images are Level 1+ products, in which the dark current and flat field have been corrected, and a speckle masking method was used to calibrate the imaging quality \citep{2016NewA...49....8X}. NVST H$\alpha$ line-center and the H$\alpha$ off-band images were first co-aligned with each other using the method of \citet{2024ApJ...977..186C}, and then registered to a standard calibrated solar full disk via an automatic mapping approach \citep{2019AnJi}.

IRIS tracked the AR 12782 in a large, dense 320-step raster, resulting in a field of view \( 167\ ^{\prime\prime} \times 175\ ^{\prime\prime} \). The Slit-jaw images (SJIs) observed by IRIS at Si~{\sc{iv}} 1400 \AA, 1330 \AA, and Mg~{\sc{ii}} k 2796 \AA~were used. Their temporal cadence is 37 s and spatial resolution of \(0''.35\, \text{pixel}^{-1} \). 
The calibrated Level-2 IRIS spectral data were used, which have already been processed with dark current subtraction, flat fielding, orbital variation, and geometrical corrections \citep{2014SoPh..289.2733D}. To characterize the mass flows in the forming filament and associated local heating at the filament footpoints, the following spectral lines were used in this study: (1) Si~{\sc{iv}} 1394/1403 \AA\ lines, formed in the transition region (TR) with typical formation temperatures of around 0.08 MK and 0.15 MK, respectively; (2) Mg~{\sc{ii}} k 2796 \AA\ line, with a formation temperature of 0.01 MK in the chromosphere; and (3) C~{\sc{ii}} 1334/1335 \AA\ lines, with a formation temperature of around 0.25 MK in the upper chromosphere or lower TR. We applied a single Gaussian fit to all the optically thin Si~{\sc{iv}} 1403 Å line profiles along the slit to determine the line parameters, including peak intensity, line width, and Doppler shift.
Spectral lines from nearby neutral and singly ionized atoms in a quiet-Sun region were used for absolute wavelength calibration \citep[e.g.,][]{2016ApJ...824...96T,2022A&A...659A.107C}.

Additionally, extreme ultraviolet (EUV) 304 \AA\ imaging data, with a spatial resolution of  \(0''.6\, \text{pixel}^{-1} \)  and a temporal cadence of 12 s, provided by the Atmospheric Imaging Assembly \citep[AIA,][]{2012SoPh..275...17L}, and line-of-sight 45 s magnetograms, with a spatial resolution of \( 0''.5 \, \text{pixel}^{-1} \), from the Helioseismic and Magnetic Imager \citep[HMI,][]{2012SoPh..275..207S} onboard the Solar Dynamics Observatory (SDO) \citep[SDO,][]{2012SoPh..275....3P} were also used. To compensate for solar rotation, all the SDO, NVST, and IRIS/SJI images taken at different times were aligned to an appropriate reference time of 01:35 UT.

\section{Results} \label{sec:floats}

\begin{figure}[ht!]
\plotone{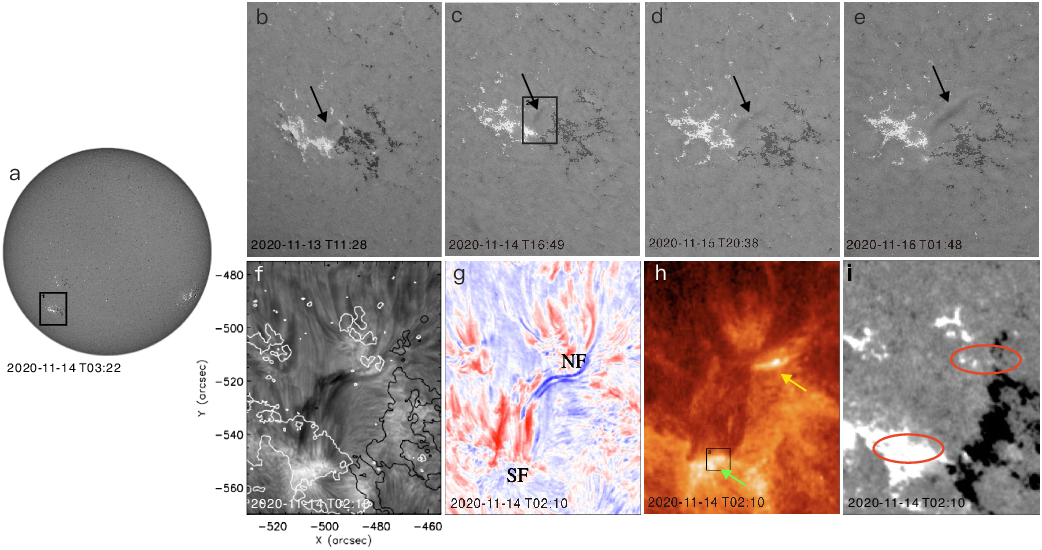}
\caption{Overview of the target filament. Panels (a)$-$(e) were composite images of GONG H$\alpha$ and SDO/HMI line-of-sight magnetograms. The black box in panel (a) marks a smaller field of view (FOV) containing the target filament of panels (b)$-$(e).  Panels (b)$-$(e) show the forming filament observed on November 13, 14, 15, and 16, respectively.  Black arrows indicate the forming filament. The black box in panel (c) provides the FOV of panels (f)$-$(i).  Panels (f) and (g) show NVST H$\alpha$ line center and Doppler proxy map constructed from the H$\alpha$ red and blue wing images. Panels (h) and (i) present the simultaneous AIA 304 \AA\ image and HMI line-of-sight  magnetogram. The simultaneous HMI line-of-sight magnetogram, with the level of  ± 100 G, was overplotted in panel (f).  Yellow and green arrows indicate the footpoint regions of the filament. Red ellipses denote their magnetic environment. An animation of panels (f)$-$(i), with a duration of 15 s, is available.}
\label{fig1}
\end{figure}

Figure \ref{fig1} depicts the formation and subsequent growth of this target AR filament. From November 14 through November 16, 2020, this filament started to form and evolved into a typical AR filament.  Note that on November 13, the target presented as a collection of chromospheric dark threads exhibiting characteristics akin to an arch filament system, overlying the PIL at a relatively large angle. By November 15, it had gradually evolved into a typical filament, aligned along the PIL with a discernible increase in its apparent length exceeding 30 Mm. Thus, we termed it as a nascent filament in the current study.

With the multi-passband  H$\alpha$ monitor of NVST on November 14,  this forming filament was found to be highly dynamic structure, consisting of relatively long dark threads (see Figure \ref{fig1} (f)).  As illustrated in Figures \ref{fig1} (g) and  \ref{fig2} (a)$-$(b), distinct upflows and downflows coexist along the filament axis. These mass flows well traced the magnetic field morphology of this forming filament, particularly its footpoints. As indicated by the green and yellow arrows in Figure \ref{fig1} (g), the south footpoints (SF) of this forming filament were rooted in the positive-polarity magnetic field region of the AR, and the north footpoints (NF) were located in the negative-polarity magnetic field flux of the AR (also see  red ellipses in Figure \ref{fig1} (i)). In accordance with the chirality definition proposed by \citep{1998SoPh..182..107M}, it is categorized as a sinistral filament. 

Based on its footpoint regions (SF and NF), it can also be noted that the forming filament appears to be a high-lying structure, with its main axis arching at a relatively large angle (about 30 degree) to the PIL of the decayed AR. This apparent angle of the filament relative to the PIL is mainly due to a prominent projection effect on November 14. Moreovor, the filament might still be in a nascent state, with its threads gradually aligning along the PIL. After careful inspection of the associated movie, no obvious helical or twisted trajectory of mass flows was discerned within this forming filament, suggesting a possible sheared arcade magnetic configuration. Such sheared arcade magnetic configurations appear to be commonly found in forming AR filaments or re-forming intermediate filaments, as suggested by previous studies {\citep[e.g.,][]{2017ApJ...835...94O,2018ApJ...869...78C}.

\begin{figure}[ht!]
\plotone{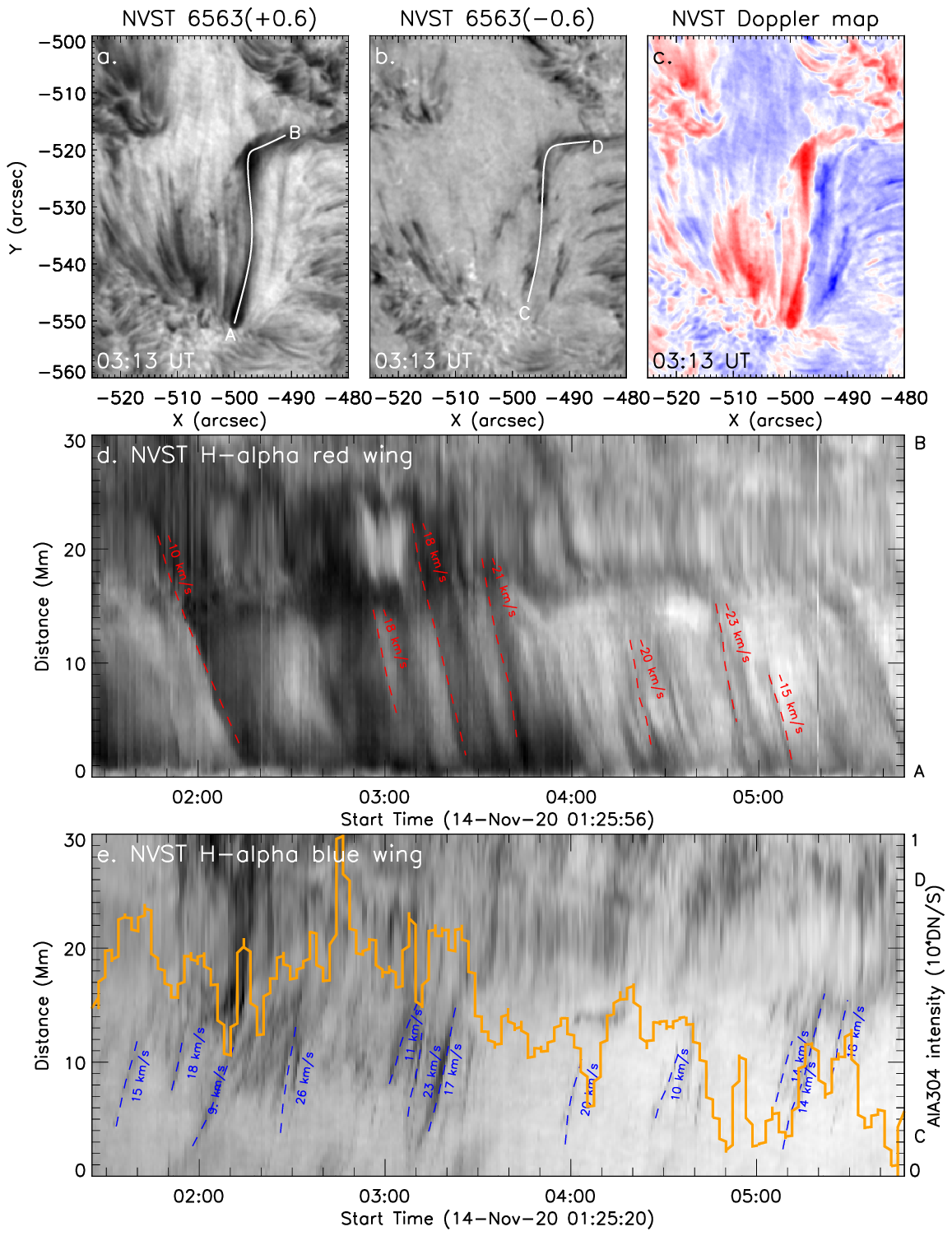}
\caption{The mass flows in the forming filament revealed by the NVST observations. (a) the red wing image of H$\alpha$ at 03:13 UT. (b) the blue wing image of H$\alpha$ at 03:13 UT. (c) the Doppler proxy map that combines (a) and (b) in their corresponding colors. Panel (d) shows the time–distance plot along the path AB, indicated in panel (a), with a downward average velocity of approximately 18 km s$^{-1}$. Panel (e) shows the time–distance plot along the path CD indicated in panel (b), with an upward average velocity of approximately 16 km s$^{-1}$. Meanwhile, the normalized 304 \AA\ EUV intensity flux at the south footpoint is overplotted as an orange curve in Panel (e).  An animation of this figure, available in the online journal, demonstrates the counterstreaming flows observed by the NVST. The animation's duration is 15 s.}
\label{fig2}
\end{figure}

As shown in Figure \ref{fig1} (g) and associated movie, persistent blueshift and redshift coexisted along the filament's main axis. At certain times, blueshift and redshift overlapped or appeared at different ends of the filament (see Figure \ref{fig1} (g),  while at other times they appeared along adjacent filament threads (see Figure \ref{fig2} (c)). These signals indicate the occurrence of filament counterstreaming flows, a long-lived phenomenon that persisted throughout the formation and evolution of this filament from 01:25 UT to 05:45 UT. To characterize these H$\alpha$ component upflows and downflows, two time-distance plots were generated along slices AB and CD in Figure \ref{fig2} (a) and (b). The results are presented in Figure \ref{fig2} (d) and (e), respectively. In the H$\alpha$ red wing (6563.6 \AA), significant downflows, with a mean projected velocities ranging from -23 to -10 km s$^{-1}$, drained from the NF to the SF. In the H$\alpha$ blue wing (6562.4 \AA), upflows with velocities ranging from 9 to 26 km s$^{-1}$ were observed, streaming from SF to NF. The most striking feature observed here is the presence of upward mass flows only in the H$\alpha$ blue wing and downward mass flows only in the H$\alpha$ red wing. This result strongly supports the view that counterstreaming flows in this forming filament represent interleaved unidirectional mass flows in opposing directions along neighboring filament threads, rather than filament mass longitudinal oscillation.  As shown in Table 1, the mean velocity of these downflows/upflows detected in Figure \ref{fig2} is -18 km s$^{-1}$ and 16 km s$^{-1}$, respectively. The velocities of these downflows and upflows fall within the typical velocity range of previously observed H$\alpha$ counterstreaming flows  \citep[e.g.,][]{1998Natur.396..440Z,2005SoPh..226..239L}.

Figure \ref{fig3} shows the IRIS spectroscopic observation of this forming filament during 02:45-03:34 UT.  As illustrated in Figure \ref{fig3} (c), the ongoing counterstreaming flows are also evident in the Doppler shift map of the Si~{\sc{iv}} 1402.77 \AA\ line. Its global Doppler shift pattern well aligns with that revealed in the NVST H$\alpha$ Doppler proxy map. The absolute Doppler shift velocities of these Si~{\sc{iv}} counterstreaming flows range from 20 to 45 km s$^{-1}$, which is significantly higher than the flow velocities observed in H$\alpha$ passbands.  
Meanwhile, Figure \ref{fig3} (b) reveals a greatly enhanced line width velocity, with a typical value of 40 km s$^{-1}$, within the forming filament. Given a thermal broadening velocity of 6.8 km s$^{-1}$ and an instrumental broadening  velocity of 4.1 km s$^{-1}$ \citep[e.g.,][]{2014Sci...346A.315T,2017ApJ...838....2Z}, the nonthermal width velocity in the forming filament is of the same order. It may be caused by the significant micro-turbulence generated by anti-paralleled mass flows.  Furthermore, using the wavelength scan of the Mg ~{\sc{ii}} k 2796.35 \AA\ line, Figure \ref{fig3} (e)-(h) reveals the fine-scale motion structures of the forming filament across different velocity ranges spanning the lower to upper chromosphere. In the red wing of the Mg~{\sc{ii}} k line, evident mass flows drain downward along the forming filament toward its south end (as marked by the white arrow pair in Figure \ref{fig3} (g)). The redshifted velocity ranges from -20 to -60 km s$^{-1}$. In the blue wing of the Mg ~{\sc{ii}} k line,  sporadic and threadlike upflows emanate from the filament's south end (as marked by the yellow arrows in Figure \ref{fig3} (e)), in the blueshifted velocity range of 32 to 70 km s$^{-1}$. Combined with the NVST observations, this result indicates counterstreaming flows in the forming filament occur in both chromospheric and transition region temperatures. 

The peak intensity image of the Si~{\sc{iv}} 1402.77 \AA\ line in  Figure \ref{fig3} (a) reveals a series of compact UV brightenings appearing at the SF and NF regions of this forming filament (as denoted by black and white circles}). The IRIS slit crossed some of these compact UV brightenings. Figure \ref{fig3} (i)-(k) presents the spectral profiles of the Si ~{\sc{iv}} 1402.77 \AA\ line, the C ~{\sc{ii}} line pair (1334.5 \AA\ and 1335.7 \AA), and the Si ~{\sc{iv}} 1394 \AA\ line extracted from the NF and SF regions of the filament (see red and blue boxes in circles). Compared with the reference spectra extracted from a nearby plage region (indicated by the black dashed-line box in Figure \ref{fig3} (a) and (e)), spectral lines in the SF and  NF regions reveal a clear enhancement and evident line broadening in both Si~{\sc{iv}} and C~{\sc{ii}} lines, while still maintaining a Gaussian profile. Notably, the C~{\sc{ii}} line pair in the NF region shows a reversal at the line core, which is absent in the SF region. Moreover, similar compact footpoint brightenings were also detected in AIA EUV 304 Å observations (Figure \ref{fig1}h and the associated movie). In Figure \ref{fig2}e, the normalized 304 \AA~AIA flux, calculated within the SF region (denoted by the box in Figure \ref{fig1}h)), indicates a series of repeated enhanced flux peaks. These repeated EUV brightenings appear to be associated with the temporal occurrence of episodes  of H$\alpha$ upflows, repeatedly feeding mass supply into the forming filament.

Indeed, a closer look at the SF region of the filament indicates such compact brightenings associated with a series of small-scale upflows, as marked by the yellow arrows in Figures \ref{fig3}a and \ref{fig3}a. These small-scale upflows, known as network jets, were first reported by \citet{2014Sci...346A.315T} using IRIS slit-jaw imaging observations with the 1300 Å and 1400 Å filters. They are often preceded by compact network brightenings, representing short-lived, high-speed upflows, with speeds ranging from 80 to 200 km s$^{-1}$ and temperatures of at least $\sim$ 10$^5$ K. 
Figure \ref{fig4} presents one selected episode of such small-scale jets at the SF region. The multiwavelength images from NVST, IRIS, and AIA reveal that these small-scale jets exhibit  multi-temperature components (as marked by arrows), which is in line with the very recent observational results of network jets \citep{2019Sci...366..890S,2025ApJ...979..195D,2025ApJ...981..185L}. In H$\alpha$ blue wing images, these cool jet components appeared to be taller and darker upflows, resembling the enhanced spicules reported by \citet{2019Sci...366..890S} and macrospicules reported by \citet{2023ApJ...942L..22D}, while they are less continuous in AIA 304 \AA\ passbands.
The time-distance plots shown in Figure \ref{fig4} (d)-(f) clearly demonstrate that these small-scale jets are driving multi-thermal mass into the filament's end along the slice path. Their cooler H$\alpha$ jet components exhibit lower ejection velocities of 20-38 km s$^{-1}$ (with a mean of 29 km s$^{-1}$ as shown in Table 1). In contrast, the warmer UV/EUV jet components are ejected with higher velocities, ranging from 30-79 km s$^{-1}$ (with a mean of 44 km s$^{-1}$ in Table 1).  Compared to the H$\alpha$ upflows detected along the filament spine (which have a mean velocity of 16 km s$^{-1}$), the cool jet components newly initiated at the filament's end exhibit a larger velocity and possibly no deceleration. 

\begin{table}[ht]
\label{tab:}
\caption{Detailed information of the upflows/downflows detected at different imaging passbands}
\centering
\begin{tabular}{cccccc} 
\toprule
Mass flows & Event &Star time &  Duration & Velocity & Band \\ & No. & (UT) & (min) & (km/s) & (Å) \\ 
\hline
 & 1 & 01:50 & 25 & -10 & \\
 & 2 & 02:55 & 10 & -18 & \\
 & 3 & 03:10 & 15 & -18 & \\
Downflows along the filament spine & 4 & 03:32 & 11 & -21 & NVST H$\alpha$ +0.6\\
  detected in Figure \ref{fig2}(d)& 5 & 04:20 & 6 & -20 &  \\
 & 6 & 04:50 & 5 & -23 & \\
 & 7 & 05:03 & 8 & -15 & \\
\hline
 & 1 & 01:32 & 8 & 15 & \\
 & 2 & 01:50 & 6 & 18 & \\
 & 3 & 01:56 & 15 & 9 & \\
 & 4 & 02:26 & 5 & 26 & \\
 & 5 & 03:00 & 10 & 11 & \\
Upflows along the filament spine  & 6 & 03:08 & 6 & 23 & NVST/H$\alpha$ -0.6\\
 detected in Figure \ref{fig2}(e)& 7 & 03:14 & 9 & 17 & \\
 & 8 & 03:59 & 4 & 20 & \\
 & 9 & 04:26 & 9 & 10 & \\
 & 10 & 05:05 & 7 & 14 & \\
 & 11 & 05:07 & 16 & 14 & \\
 & 12 & 05:25 & 5 & 18 & \\
\hline
 & 1 & 03:09 & $>$3 & 38 & \\
 & 2 & 03:13 & $>$2 & 20 & \\
Upflows at the filament's end & 3 & 03:14 & $>$2 & 27 & NVST/H$\alpha$ -0.6\\
detected in Figure \ref{fig4}(d)  & 4 & 03:17 & $>$2 & 34 & \\
 & 5 & 03:18 & $>$2 & 26 & \\
\hline
 & 1 & 03:04 & 2 & 29 & \\
 & 2 & 03:10 & 1 & 44 & \\
Upflows at the filament's end & 3 & 03:11 & 2 & 14 & SDO/AIA 304\\
 detected in Figure \ref{fig4}(e) & 4 & 03:17 & 0.5 & 69 & \\
 & 5 & 03:18 & 3 & 27 & \\
 & 6 & 03:25 & 1 & 59 & \\
\hline
Upflows at the filament's end & 1 & 03:10 & 1 & 79 & IRIS/SJI 1400\\
 detected in Figure \ref{fig4}(f) & 2 & 03:23 & 1 & 34 & \\
 \hline
\end{tabular}
\tablecomments{Duration: the time period spanning from the initial appearance to the final disappearance of the upflows or downflows, determined from  the corresponding time-distance plots. The duration of the H$\alpha$ -0.6 upflows in Figure 4(d) should be considered lower limits, due to the shorter slice path of the time-distance plots. Velocity: upward flows are assigned a positive value, while downward flows are assigned a negative value. Along the filament spine, downflows and upflows in H$\alpha$ passbands have a mean velocity of -18 km s$^{-1}$ and 16 km s$^{-1}$, respectively.  At the filament's south end, upflows have a higher mean velocities:  29 km s$^{-1}$ in H$\alpha$ passband and 44 km s$^{-1}$ in EUV/EUV passbands.}
\end{table}

\begin{figure}[ht!]
\plotone{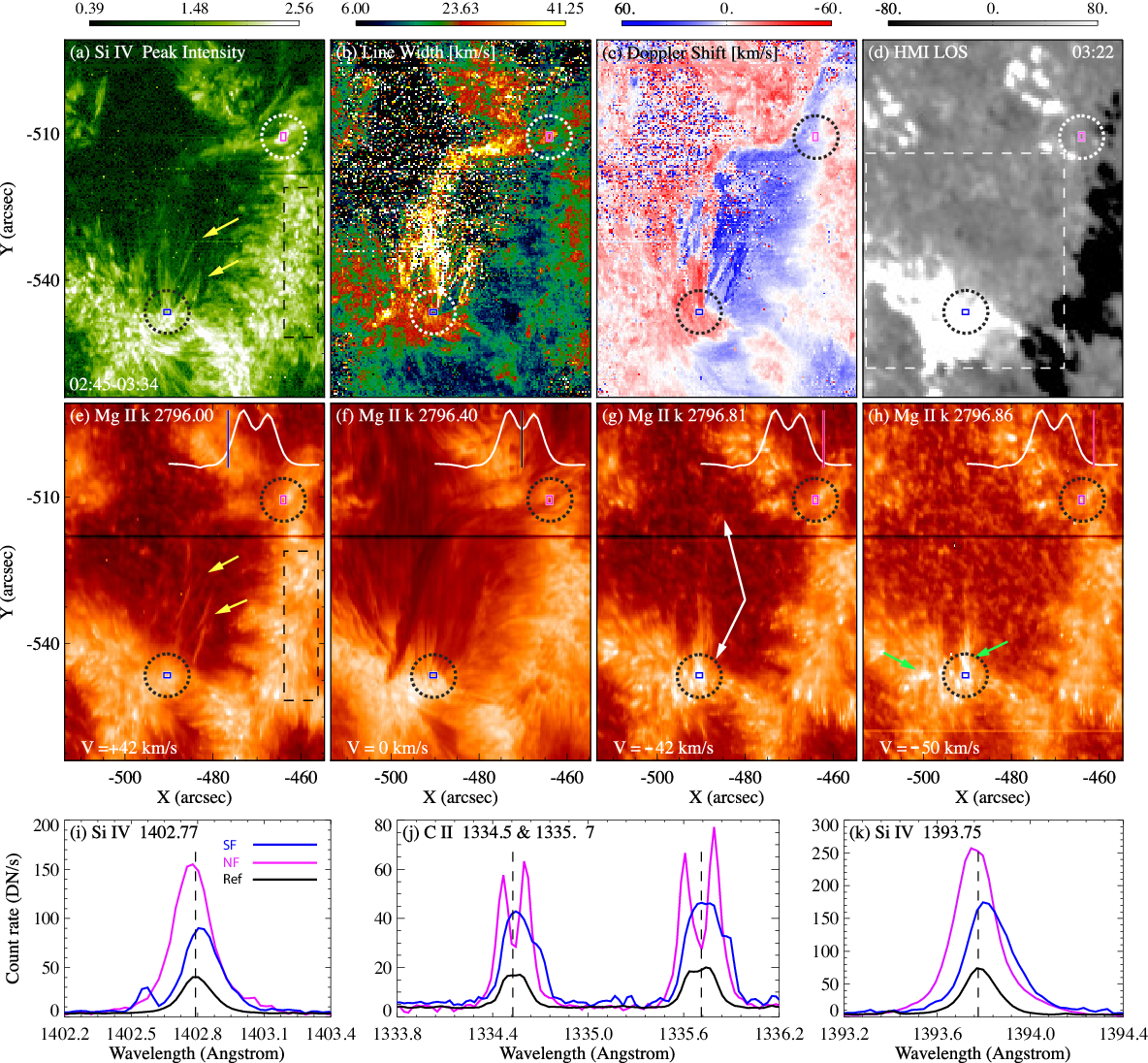}
\caption{Interface Region Imaging Spectrograph (IRIS) spectroscopic observations of the forming filament during 02:45 to 03:34 UT. (a-c): The peak intensity, line width, and Doppler shift of the Si ~{\sc{iv}} line at 1393.75 \AA, obtained by a single Gaussian fit. (d) The HMI LOS magnetogram, with the level of ±80 G, at 03:22 UT. The white dashed-line box denotes the FOV of Figure \ref{fig6} (a). (e)$-$(h): Snapshots of Mg~{\sc{ii}} k line at different wavelength from blue wing to red wing, showing the mass flows along the forming filament at different velocities. Blue boxes in Figures (a) to (h) mark the areas where the spectral lines at the southern footpoint of the filament are extracted and spatially-averaged, and the purple boxes mark the areas at the northern footpoint. It is marked with a more obvious black and white dotted circle in the picture. The black dashed boxes in (a) and (e) are the selected reference patches.
 (i)$-$(k): Spectral lines of Si~{\sc{iv}} 1402.77 \AA, C~{\sc{ii}} 1334.5 \AA~ and 1335.7 \AA, and Si  ~{\sc{iv}} 1394 \AA~at the south footpoint region (blue box), the north footpoint region (purple box) of the forming filament, and the reference plage region (black dashed-line box), respectively. Yellow arrows in panels (a) and (e) indicate numerous small-scale jets, representing thread-like upflows along the forming filament. The white arrow pair in panel (g) indicates mass flows within the forming filament, with a distinct brightening region at its south footpoint. Green arrows in panel (h) indicate brightenings in the low chromosphere.
An animation demonstrates the counterstreaming flows and footpoint region brightenings revealed by the IRIS wavelength scan of the Mg~{\sc{ii}} k line. The animation's duration is 14 s.
\label{fig3}}
\end{figure}


\begin{figure}[ht!]
\plotone{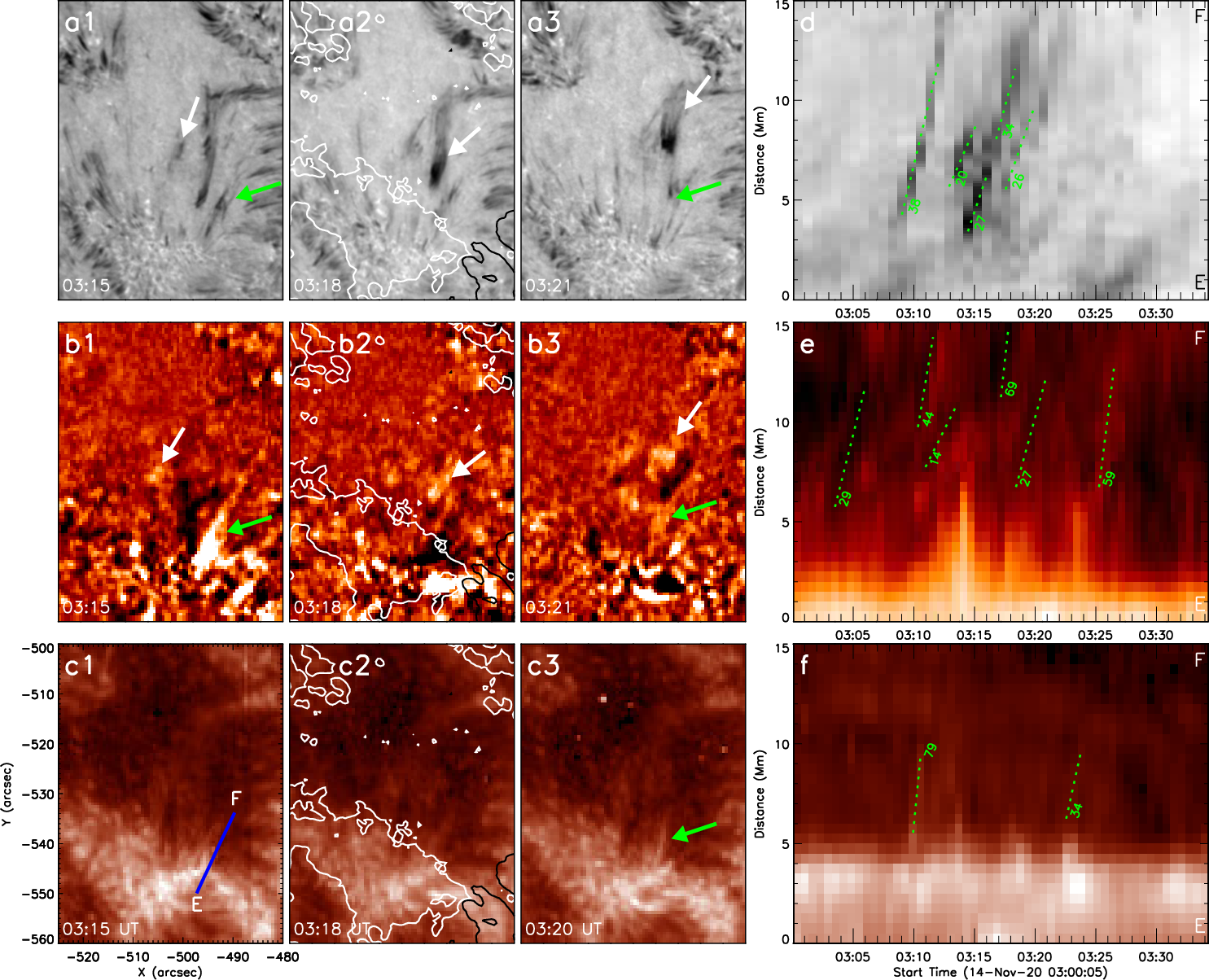}
\caption{Multi-temperature nature of a selected episode of small-scale jets observed at the south footpoint of the forming filament. Panels  (a), (b), and (c) give the NVST H$\alpha$ blue wing images, AIA 304 \AA~images, and IRIS/SJI 1400 \AA~images, respectively.  HMI contours of level ±30 G are overlaid in the second column. The green and white arrows mark multi-temperature components of several ongoing small-scale jets. (e)$-$(f): Time-distance plots of NVST H$\alpha$ blue wing, AIA 304 \AA, and IRIS/SJI 1400 \AA~images along the slice EF in (c1). An animation of this figure is available in the online journal, showing the occurrence of small-scale jets. The duration of the animation is 3 s.}
\label{fig4}
\end{figure}

\begin{figure}[ht!]
\centering
\includegraphics[scale=0.6]{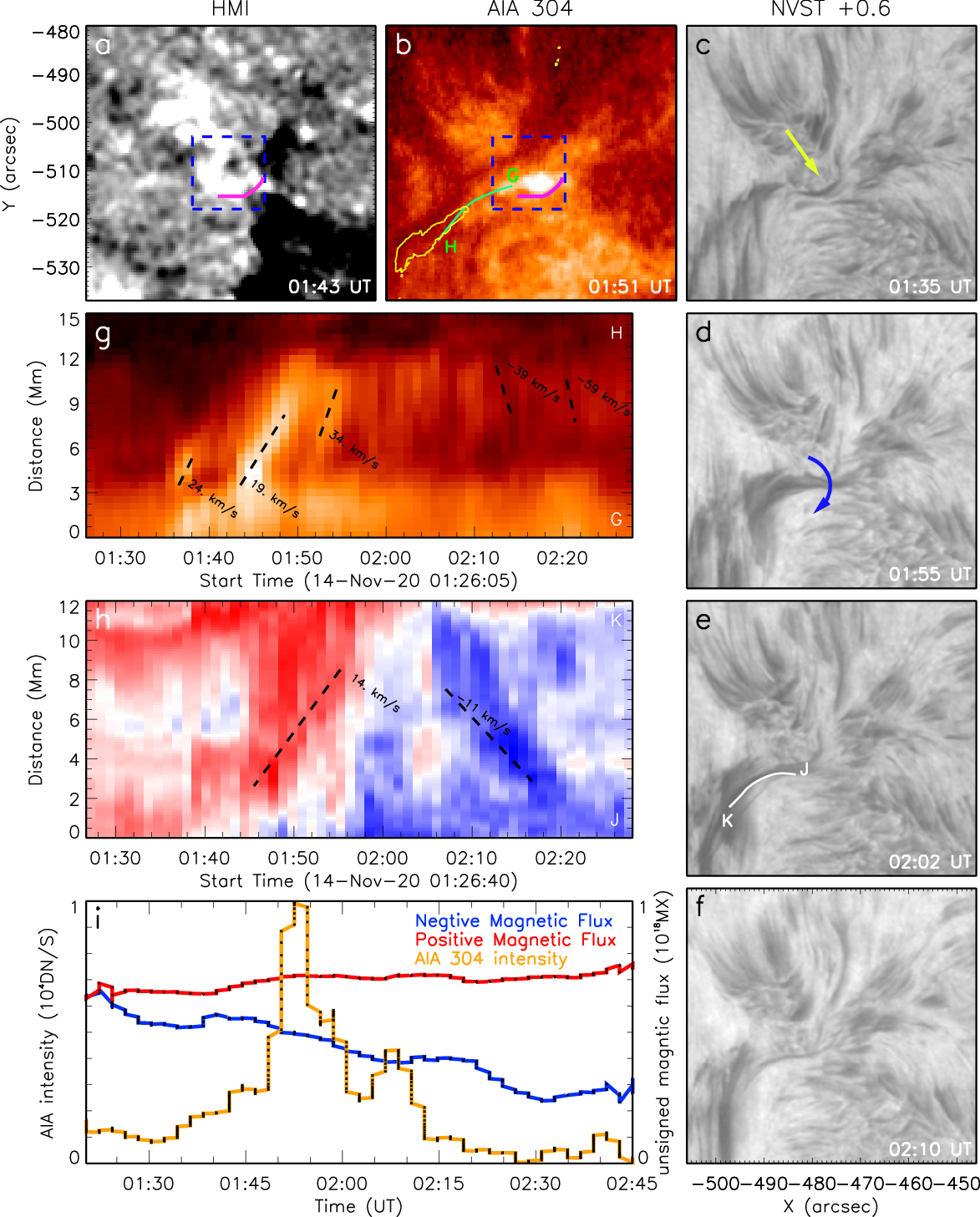}
\caption{Small-scale jets and associated minifilament eruption observed at the north footpoint of the forming filament. Panels (a) and (b) show SDO/HMI line-of-sight magnetogram and AIA 304 \AA~image near the northern footpoint of the filament. The blue dashed box shows the area selected for the calculation of the magnetic flux. The yellow contour represents the outline of the forming filament observed in the NVST H$\alpha$ line core. Panels (c)–(f) show NVST H$\alpha$ red wing images of the same FOV, demonstrating the minifilament eruption and associated untwisting motion. Panel (g) is the time-distance plot along the path GH in panel (b), showing upward and downward flows. Panel (h) presents the time-distance plot derived from Doppler proxy images along the spatial path JK indicated in panel (e), wherein signals corresponding to both the red and blue wings of the spectral line are displayed. Panel (i) shows the magnetic flux variations and the normalized radiative intensity changes of AIA 304 \AA. An animation shows the minifilament eruption and magnetic flux variation in its source region. The duration of the animation is 2 s.}
\label{fig5}
\end{figure}

\begin{figure}[ht!]
\plotone{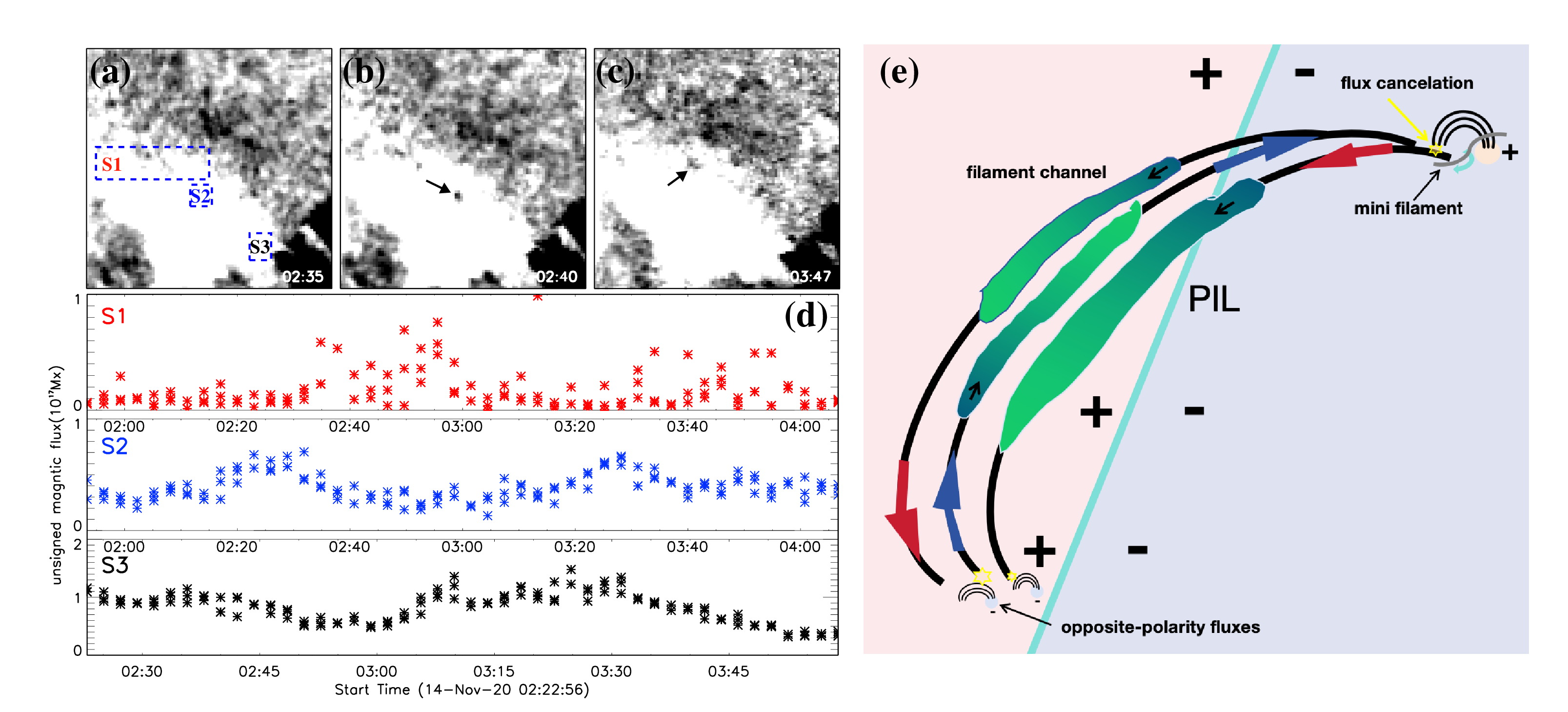}
\caption{(a)–(c) Snapshots of selected HMI LOS magnetograms with a magnetic level of ±20 G. In panel (a), three small regions, S1, S2, and S3, near the southern footpoint were selected to calculate their magnetic flux. (d) The magnetic flux evolution in these three regions. (e) A schematic diagram illustrating the proposed scenario, explaining the nature and origin of the forming filament based on observations. An animation is available, showing the occurrence of emergence and disappearance of small-scale negative-polarity magnetic flux elements. The duration of the animation is 5 s.}
\label{fig6}
\end{figure}


Figure \ref{fig5} presents the EUV brightening and associated small-scale jets at the NF region. As shown in Figure \ref{fig5} (a) and (b), the EUV brightening at the NF region occurred below the north end of the forming filament, located at a magnetic mixed-polarity region. At this location, small-scale opposite-polarity magnetic fluxes converged and canceled with each other, which results in an obvious decrease of negative flux in its source region (see Figure \ref{fig5} (i)).  Using the NVST red wing observations, we identified a minifilament, with a projected length of approximately 10 Mm, at the same location prior to the EUV brightening at 01:35 UT. The minifilament began to erupt at approximately 01:40 UT. Its eruption then led to the EUV brightening in the AIA 304 \AA\ passband and an obvious rolling motion along the long thread of the forming filament in the NVST red-wing images (as indicated by the blue circular arrow in Figure \ref{fig5} (d). Subsequently, a small-scale jet occurred following this rolling motion, injecting mass into the forming filament.
Such apparent rolling motions are most likely triggered by an rapid twist transfer from the eruption minifilament into the forming filament. Similar phenomena have been reported in the filament eruptions in the footpoints of coronal loops \citep[e.g.,][]{2019ApJ...887..239Y,2019FrASS...6...44L,2021ApJ...911...33C,2024ApJ...974L...3Z}. 
To characterize the mass motion associated with this jet, two time-distance plots were generated along slices GH and JK, respectively.
As shown in Figure \ref{fig5} (g), this small-scale jet manifested as bright EUV upflows, with measured velocities of 19$-$34 km s$^{-1}$, in the AIA 304 \AA\ passband. Subsequently, anti-parallel dark EUV downflows drained out along the same path, with velocities of 39$-$59 km s$^{-1}$. In the H$\alpha$ passband,  the jet appeared as cool mass upflows and downflows, with velocities ranging from 11 to 14 km s$^{-1}$. It is important to note that the H$\alpha$ upflows/downflows were respectively detected in the red/blue wing passbands due to the projection effect in the southern hemisphere. 

Flux cancellation refers the mutual close and then disappearance of opposite-polarity magnetic fragments in the photosphere, which has been suggested to be signatures of magnetic reconnection at lower atmosphere  \citep[][and its references]{1987ARA&A..25...83Z}.
Extensive observations reveal that flux cancellation are closely related to the formation and eruption of minifilaments \citep[e.g.,][]{2011ApJ...738L..20H,2020ApJ...902....8C,2025ApJ...983..143C} and the solar jet activities, i.e., coronal jets \citep[e.g.,][]{2016ApJ...821..100S,2017ApJ...844...28S,2018ApJ...853..189P,2023ApJ...942...86Y}, network jets/jetlets \citep[e.g.,][]{2020ApJ...897L...2P,2025ApJ...979..195D,2025arXiv250510995Y},  as well as enhanced chromospheric spicule activities \citep{2019Sci...366..890S,2023ApJ...942L..22D,2025ApJ...985L..47B}.
From Figure 5 (g) and (h), it is evident that the minifilament-induced EUV brightening and mass flows both occurred at the north filament's end during the period 01:45 - 02:00 UT.  Obvious flux cancellation occur below the minifilament site. In Figure \ref{fig5} (i), the AIA 304 \AA\ flux and magnetic flux variation were calculated within the source region of the erupted minifilament (as indicated by the blue dashed box in Figure \ref{fig5} (a) and (c)). Before 01:40 UT in the source region of filament, both the positive and negative magnetic fluxes were decreasing. After that, the negative flux decreased significantly, while the positive flux increased slightly. This process indicates that the cancellation of the magnetic flux occurred, with a magnitude of 10$^{17}$ Mx.
Similar flux cancellation at smaller scales also occur at the filament's south end. Figure \ref{fig6} (a)-(c) and associated movie present the photospheric magnetic evolution at the SF region. In fact, small-scale jets at SF region originated from a positive-polarity magnetic field region, but in which numerous very weak, small-scale negative-polarity flux elements randomly emerged. Such random negative-polarity flux emergences are only noticeable when the magnetic level of the HMI LOS magnetograms set as $\pm$ 20 G. As marked by the black arrows in Figure \ref{fig6} (b) and (c), their spatial scale ranges from approximately 2 to 5 pixels, namely 0.7 to 1.8 Mm, which approaches the spatial resolution limit of the HMI. At the SF region, these small-scale flux elements rapidly emerged and soon disappeared within several minutes, occurring near the starting points of the small-scale jets and blueshift sources. Figure \ref{fig6} (d) shows the flux variation of negative-polarity elements in three selected box regions (S1, S2, and S3 denoted as blue dashed boxes in Figure \ref{fig6} (a) ). The magnetic flux variation is on the order of 10$^{17}$ Mx, indicating a very weak flux emergence and possible subsequent cancellation. Consistent with the views mentioned above, these results suggest that flux cancellation may play an important role in triggering these small-scale jets at the filament's ends.

\section{Discussion} \label{sec:cite}
Counterstreaming flows are a prevalent intrinsic motion within filaments, and understanding their observational characteristics and physical origins is a fundamental problem. Over the past decades, extensive solar imaging observations have revealed the frequent occurrence of jet activity along the legs of coronal loops or at filament footpoints \citep[e.g.,][]{2015ApJ...815...71C,2016ApJ...821..100S,2017ApJ...835...35H,2019ApJ...887..239Y,2019MNRAS.488.3794W}. Consequently, some researchers proposed that jet-like mass ejections may be the direct cause of counterstreaming flows in filaments \citep{1999ApJ...520L..71W,2000SoPh..195..333C}; however, this hypothesis has hitherto lacked robust observational supports \citep[see][]{2010SSRv..151..333M}.  Recently, \citet{2025ApJ...987...38Z} reported that recurrent coronal jets  at one end of a solar prominence induced both forward- and backward-moving mass flows along the adjacent magnetic field lines of the prominence cavity, consequently giving rise to a Doppler Bullseye Pattern within the cavity as observed from the limb.
Recent observations by \citet{2020ApJ...897L...2P} provided solid evidence that recurrent network jet eruptions directly drive EUV counterstreaming flows along filament/filament channel threads. Compared with their observations, our current findings further reveal that: (1) counterstreaming flows along the threads of a forming filament/filament channel comprise plasma ranging from chromospheric to transition region temperatures; and (2) the cool and warm components of the counterstreaming flows are directly driven by a series of small-scale jets emanating from chromospheric and transition region heights; (3) UV/EUV brightenings near the SF and NF regions, as revealed by IRIS spectroscopy, indicate local heating at the filament footpoints. Regarding the origin of these small-scale jets, our observations also support the view that flux cancellation results in the occurrence of small-scale jets at both the SF and NF regions. Specifically, for the jets observed in the NF region, high-resolution chromospheric observations from NVST revealed the presence and subsequent eruption of a small minifilament after flux cancellation. Consistent with previous observations \citep[e.g.;][]{2018ApJ...853..189P,2020ApJ...902....8C}, this result suggests that flux cancellation may be the primary candidate for triggering the minifilament eruption and its associated jet activity. This eruption not only injected its mass but also transferred its magnetic twist into the developing active region filament. 
A notable feature of the current study is the occurrence of small-scale jets in the SF region within a predominantly positive-polarity plage region. At the detection limit of the HMI magnetic field resolution, we identified numerous small-scale negative-polarity flux elements emerging and disappearing with a magnitude of $\sim$ 10$^{17}$ Mx. This implies that flux emergence and cancellation occurring near or even below the resolution limit of HMI may be present in the base region of these small-scale jets. Therefore, we conjecture that smaller-scale opposite-polarity flux elements and its associated flux cancellation processes may drive these small-scale jets observed in the SF region.

Based on our results described above, a possible scenario was proposed in Figure \ref{fig6} (e) to interpret the current observations. The forming filament arcade above the PIL, with its south footpoints anchored at a predominantly positive-polarity plage region. The emergence and subsequent disappearance of negative-polarity minor magnetic polarities may trigger repeated magnetic reconnection events (illustrated as a small arch in Figure \ref{fig6} (e)), resulting in a series of small-scale jets and UV/EUV compact brightenings. These recurrent jets inject multi-temperature plasma upward into the filament, continuously supplying the necessary material and energy to sustain the observed counterstreaming flows. At the filament's northern footpoint, magnetic flux converges and cancels in the photosphere. This process lead to the formation and eruption of a tiny minifilament.  Its eruption undergoes unwinding, injecting both twist and material into the forming filament. Due to the occurrence of recurrent small-scale jets and associated local heating, continueous mass flows form along different nearby filament threads. Notably,  these mass flows, characterized as Doppler red/blue-shifts, occur sequentially along individual magnetic field lines and coexist simultaneously along different field lines within the filament, eventually causing the apparent counterstreaming flows.

Local heating at the footpoints of magnetic fields is a crucial input in MHD numerical simulations of solar prominences/filaments and coronal rain \citep[e.g.,][]{2012ApJ...748L..26X,2016ApJ...823...22X,2023A&A...670A..64J}, but observational constraints on the occurrence and nature of this local heating are currently lacking.  Within the chromospheric evaporation-coronal condensation framework, researchers have typically assumed localized random heating at the footpoints of magnetic arcades, distributed randomly both spatially and temporally, with the heating being one to three orders of magnitude stronger than the background heating \citep[e.g.,][]{2011ApJ...737...27X,2012ApJ...748L..26X,2020NatAs...4..994Z,2025ApJ...987...38Z}. Using high-resolution imaging in He~{\sc{i}} 10830 \AA\ and H$\alpha$ observations of a set of active region (AR) loops associated with counterstreaming motions, \citet{2019ApJ...881L..25Y} reported weak extreme ultraviolet (EUV) brightenings in intergranular lane areas and suggested that these subtle EUV brightenings are a possible localized heating signal at the footpoints of AR loops. \citet{2016A&A...589A.114L} reported the detection of subarcsecond UV bright points located beneath a quiescent prominence, and suggested that these brigtening features, in conjunction with associated quasi-periodic upflows, may facilitate the transport of mass into the overlying large-scale prominence.
In this work, our observations provide direct IRIS spectral and imaging evidence of local heating at the filament footpoints. At the compact brightenings in the SF and  NF regions , spectral profiles of Si~{\sc{iv}}and C~{\sc{ii}} lines reveal a clear enhancement and evident line broadening. The increase in the intensity in both Si~{\sc{iv}} and C~{\sc{ii}} lines may be attributable to localized thermal enhancement of cooler plasma within the upper chromosphere. The line broadening of the Si~{\sc{iv}} and C~{\sc{ii}} lines was possibly caused by the presence of enhanced  microturbulence generated by energy release. These results imply that unresolved reconnection processes driven by small-scale flux emergence and cancellation may be responsible for such localized heating and small-scale jet activity. Furthermore, the temporal variation of the 304 Å flux observed in the SF region (in Figure \ref{fig2} (e)) indicates that the local footpoint heating could be a random and recurrent phenomenon, possibly modulated by turbulent photospheric motions. The advent of new solar instruments, including the Daniel K. Inouye Solar Telescope (DKIST) and the Chinese Giant Solar Telescope (CGST), will provide higher resolution imaging and spectroscopic observations, enabling us to confirm this possibility in the future.

\section{Summary } \label{sec:cite}
Utilizing multi-wavelength, high-resolution imaging and spectroscopic observations from NVST, SDO, and IRIS, we examined the nature and origin of counterstreaming flows within a forming AR filament. This work elucidates the multi-thermal origin of counterstreaming flows within a forming filament and provides evidence of localized chromospheric heating at the filament footpoints. Our main findings are listed as follows:
\begin{itemize}
    \item The forming filament exhibits long-lived and multi-thermal counterstreaming flows, persisting for at least several hours in our observations. Our results reveal that these flows are primarily interleaved unidirectional mass flows in opposing directions. In the H$\alpha$ passband, their typical speeds are around 20 km s$^{-1}$, while in the UV Si~{\sc{iv}} 1400 \AA~and EUV 304 \AA~passbands, their flow speeds can reach up to 40 km s$^{-1}$. IRIS spectral diagnostics show enhanced micro-turbulence in the filament counter-streaming flow regions, as indicated by the significant nonthermal width velocity of $\sim$ 40 km s$^{-1}$ in Si~{\sc{iv}} 1400 Å spectral profiles.
    \item  Imaging observation reveals these multi-thermal flows emanating from small-scale brightenings at the filament's ends. At their nascent phase, these flows manifest as a series of small-scale, collimated small-scale jets at the filament's ends. They continuously inject cool chromospheric and warm transition region plasma into the filament channel, hereby continuously supplying multi-temperature plasma into the filament's interior and driving the counterstreaming flows. The averaged jet speeds detected at the filament's end is about  30 km s$^{-1}$ in H$\alpha$ and 44 km s$^{-1}$ in EUV/UV passbands, respectively.
    \item The small-scale brightenings appeared at the filament's ends, where the multi-thermal jets originated, IRIS spectral diagnostics confirm significant chromospheric local heating. This is indicated by pronounced intensity enhancements and line broadening of the Si~{\sc{iv}} and C~{\sc{ii}} spectral lines. 
    \item HMI observations reveal small-scale photospheric magnetic elements with opposing polarities undergoing flux emergence and/or cancellation at the filament's ends. This weak magnetic field activities suggest that unresolved magnetic reconnection events associated with flux cancellation may be driving these small-scale jets and associated localized footpoint heating.
 \end{itemize}

\begin{acknowledgments}
The authors sincerely thank the referee for the constructive suggestions/comments and Professor Chen Peng-fei from Nanjing University for the helpful discussions.
This work is supported by the NSFC grants (12325303, 12073022, 12473059, 12103005), the Yunnan Key Laboratory of Solar Physics and Space Science (YNSPCC202210, 202205AG070009), the Yunnan Provincial Basic Research Projects (202401CF070165, 202501AW070002, 202301AT070350), as well as the Graduate Research Innovation Project of Yunnan University (KC-24249880).  We would like to thank the NVST, IRIS, and SDO teams for high-cadence-data support. 
\end{acknowledgments}

%





\bibliography{sample631}{}
\bibliographystyle{aasjournal}



\end{document}